\documentclass[amsmath,amssymb,sort&compress,citesort,superscriptaddress]{iopart}
\usepackage{amsbsy,bm,float}
\usepackage[normalem]{ulem}
\usepackage{iopams}
\usepackage[square,numbers]{natbib}
\usepackage{multirow}
\usepackage{graphicx}
\usepackage{dcolumn}
\usepackage{subfigure}
\usepackage{color}

\newcommand{\g}{pdf}
\begin{document}


\title{Bloch-sphere approach to correlated noise in coupled qubits}

\author{H{\aa}kon Brox$^1$, Joakim Bergli$^1$ and Yuri M. Galperin$^{1,2,3}$}%
\address{$^1$ Department of Physics, University of Oslo, PO Box 1048 Blindern, 0316 Oslo, Norway}
\address{$^1$ Centre for Advanced Study, Drammensveien 78, Oslo, Norway 0271, Oslo, Norway}
\address{$^1$ A. F. Ioffe Physico-Technical Institute of Russian Academy of Sciences, 194021 St. Petersburg, Russia}

\ead{hakon.brox@fys.uio.no}

\date{\today}

\begin{abstract}
\noindent 
By use of a generalized Bloch vector construction, we study the decoherence of a system composed of two interacting qubits
in a noisy environment.
In particular, we investigate the effects of correlations in the noise acting on distinct qubits.
Our treatment of the two-qubit system by use of the generalized Bloch vector leads to tractable analytic equations for the dynamics of the $4$-level Bloch vector
and allows for the application of geometrical concepts from the well known $2$-level Bloch sphere. 
We find that in the presence of correlated or anticorrelated noise, the rate of decoherence is very sensitive to the initial two-qubit state, as well as to 
the symmetry of the Hamiltonian. In the absence of symmetry in the Hamiltonian, correlations only weakly impact the decoherence rate.
\end{abstract}


\newcommand{\eq}{\! = \!}
\newcommand{\keq}{\!\! = \!\!}
\newcommand{\kadd}{\! + \!}

\section{Introduction}

Loss of coherence, or quantum decoherence, 
remains the most important obstacle to overcome in order to build a quantum computer.
Decoherence due to noise or to entanglement with uncontrollable degrees of freedom 
is responsible for the decay of coherent superpositions of qubit states. 
The result is irreversible loss of the quantum information required for operation of the device. 

Leading candidates for the design of a quantum computer are 
superconductor-based qubits involving Josephson junctions \cite{clarke08}, qubits based on nitrogen vacancies
in diamond \cite{gaebel_nphys,Hanson_science}, optically trapped ions \cite{iontraps} and quantum devices including
semiconductor quantum dots \cite{ladd,hanson_rmp,coish,tsyplyatyev}.
It is commonly assumed that the decoherence mainly originates from low-frequency noise with spectrum of $1/f$-type created by dynamic defects tunneling between two metastable states -- so-called two-level systems (TLSs).  In particular, in Josephson charge qubits the TLSs are formed by traps in
the amorphous material of the insulating barrier 
of the Josephson junction, and in the substrate required to fabricate the qubit.

While the effect of fluctuating TLSs on 
isolated qubits 
were extensively studied, see, e.g., references ~\cite{martinis04,martinis05,martinis10,tian,zgalperinprl,shnirman05,zgalperinmeso}, or \cite{bergli} for a review, it is clear that 
each of the tunneling 
charges present in the insulating substrate might influence 
several qubits fabricated on the same chip.
As a result, the noise acting on different qubits may be correlated, which can manifest itself in modifying the decoherence comparing to the case of
many uncorrelated sources of the same total intensity.
Specifically in charge qubits, we might have a situation where trapped fluctuators are located in the space between 
the Cooper-pair-boxes belonging to 
two different qubits
that may lead to correlated or anti-correlated noise acting on the two-qubit system.

Quantum oscillations in superconducting two-qubit systems were demonstrated already in 2003 \cite{tsai_nature}, and have recently been extended to three qubit systems \cite{dicarlo,martinis_nature}. The effect of correlated or partially correlated noise acting on two qubit systems have been studied theoretically in \cite{governale,wilhelm,averin,nori,guo,paladino_08,faoro,mastellone}. However, due to the complexity of the
Hilbert space of coupled qubits, the efforts have primarily resulted in numerical surveying of 
various situations. In addition a general theory for passive error protection in decoherence-free subspaces (DFS), subspaces of the multi-qubit Hilbert space which are immune to correlated noise, has been developed \cite{zanardi,lidar,fortunato02,fortunato,karasik} or see \cite{lidarbook} for a review. The work in this paper does not attempt to extend the general work on DFSs, but rather provide an efficient way of analyzing the decoherence for initial states that are not necessarily in the DFSs.

In this work, we present a Bloch vector treatment of the two-qubit decoherence problem, generalizing the well known formulation for a single qubit.
The 15-dimensional Bloch sphere of the two-qubit system, and the subspace of available positive definite density matrices, has a much richer mathematical structure than the corresponding construction for a single qubit \cite{goyal,siennicki,kimura}. Still the formalism is similar to the single-qubit problem, and we can take advantage of the familiar geometrical concepts developed for the treatment of two levels.

In order to illustrate the 4-level Bloch vector formulation, we consider a model problem of two coupled qubits, $Q_1$ and $Q_2$, subject to noise from two channels: $\xi_1(t)$ acting on $Q_1$ and $\xi_2(t)$ acting on $Q_2$. We are able to derive simple analytical formulas for the decoherence rate of two qubits as a function of the degree of correlation $S_c$
of the noise sources. Our results clearly demonstrate a strong impact of correlations on the decoherence rate that is very different in distinct subspaces of the two-qubit Hilbert space. 
This effect is sensitive to the symmetry of the two-qubit Hamiltonian and in the absence of symmetry, correlations in the noise are not important for the decoherence rate.

The rest of the paper is structured as follows. In Sec.~\ref{model} we define our model of two coupled qubits in a generalized environment, and discuss qualitative aspects of our model 
using a simplified picture.
The details of the initial dephasing of our system, assuming stationary fluctuators with unknown distribution of initial states, is studied in Sec.~\ref{initial}. The long-time decoherence in the weak coupling limit is studied in Sec.~\ref{long}. Finally, our results are discussed in Sec.~\ref{discussion}.

\section{Model}
\label{model}

Following \cite{averin}, the Hamiltonian describing two coupled qubits, each at zero bias, where the qubits are coupled directly by interaction 
between the basis-forming degrees of freedom, can be written as $H=H_0+H_1(t)$ where
\begin{eqnarray}
H_0&=&\sum\limits_{j=1,2}\Delta_j\sigma_x^j+\nu\sigma_z^1\sigma_z^2,\label{Hhh}\\
H_1(t) &=&\sum\limits_{j=1,2}\xi_j(t)\sigma_z^j.\label{Ham}
\end{eqnarray}
Here $\xi_{j}(t)$, $j\in\{1,2\}$ are random forces acting on the qubits, $Q_1$ and $Q_2$, respectively. The random forces might origin from the same source, from different sources or a combination, resulting in different degree of correlation in the signal. The degree of correlation is given by the expression
\begin{eqnarray}
S_c=\frac{\langle(\xi_1(t)-\bar{\xi}_1)(\xi_2(t)-\bar{\xi}_2)\rangle}{\sqrt{\langle(\xi_1(t)-\bar{\xi}_1)^2\rangle\langle(\xi_2(t)-\bar{\xi}_2)^2\rangle}},
\label{corrcoeff}
\end{eqnarray}
where the brackets indicate ensemble averaging over initial conditions and realizations of the noise process, and $\bar{\xi}_i \equiv \langle\xi_i(t)\rangle$.

The eigenstates and eigenvalues of the Hamiltonian~(\ref{Hhh}) are~\cite{averin}: 
\begin{eqnarray} \label{eigs}
\left|\psi_1\right\rangle&=&(\eta_+ +\eta_-)|0011\rangle_++(\eta_+ -\eta_-)|0110\rangle_+,\nonumber\\
\left|\psi_2\right\rangle&=&(\eta_+ -\eta_-)|0011\rangle_++(\eta_+ +\eta_-)|0110\rangle_+,\nonumber\\
\left|\psi_3\right\rangle&=&(\gamma_+ +\gamma_-)|0011\rangle_-+(\gamma_+ -\gamma_-)|0110\rangle_+,\nonumber\\
\left|\psi_4\right\rangle&=&(\gamma_+ -\gamma_-)|0011\rangle_-+(\gamma_+ -\gamma_-)|0110\rangle_+,\nonumber\\
E_1 &=&\Omega_+, \ E_2 =-\Omega_+,  \ E_3=\Omega_-,  \ E_4= - \Omega_-
\end{eqnarray}
where 
The states $\left|kl\right\rangle$, $k,l\in\{0,1\}$ are the eigenstates of the operator $\sigma_z^1\otimes\sigma_z^2$,
\begin{eqnarray*}
&& | ikjl\rangle_\pm \equiv (|ik\rangle \pm |jl \rangle)/2,  \quad  
\Delta_\pm =\Delta_1 \pm \Delta_2, \quad  \Omega_\pm = \sqrt{\Delta_\pm^2+\nu^2}, 
\\ 
 &&  \gamma_\pm=\sqrt{(1\pm \Delta_-/\Omega_-)/2}, \quad \eta_\pm=\sqrt{(1\pm \Delta_+/\Omega_+)/2}.
\end{eqnarray*}
%
Having defined our model and its parameters, we present our procedure for the study of 4-level systems by use of the Bloch-vector construction.

\subsection{Coordinate transformations}

Noise-induced decoherence leads to deviation from coherent oscillations of the states due to the Hamiltonian $H_0$.
In order to study the effect of noise, it is useful to first transform our system to the reference frame where the density matrix is stationary in the absence of noise. 
This procedure allows us to separate decoherence from rapid quantum oscillations.
The transformation is just the standard transform to the interaction picture:
\begin{equation}
\rho'=U^{\dagger}(t)\rho_S(t)U(t),\  V'(t)=U^{\dagger}(t)H_1(t)U(t)
\end{equation}
where $U(t)=e^{-(i/\hbar) H_0t}$, and $\rho_S(t)$ is the density matrix of the two-qubit system in the Schr\"{o}dinger picture. The explicit expression for $V'(t)$ is given by Eq.~(\ref{Vrot}) in~\ref{expressions}.

We proceed by specifying the initial state, $\rho'_0$, of our two-qubit system. 
Next, we perform a similarity 
transform
\begin{equation}
\rho=S^{-1}\rho'S,\quad V''(t)=S^{-1}V'(t)S,
\label{sim}
\end{equation}
where $S$ is the eigenvector matrix of the initial state $\rho'_0$.
The motivation behind the similarity transform is the following treatment in terms of coherence vectors on the
generalized Bloch ball.

For a two-qubit system, the set $\mathbb{E}_4$ of positive definite density matrices of trace one is a convex subset of the $4$-level Bloch sphere in $\mathbb{R}^{15}$, contrary to the $2$-dimensional case where the set of density matrices are equivalent with the set defined by the Bloch sphere. Specifically, the set of density matrices corresponding to pure states is only a $6$-dimensional subset of the surface of the $4$-level Bloch sphere. For more in-depth analysis of the geometry of Bloch vectors in the two-qubit system see \cite{jakobczyk}.
The set $\mathbb{E}_4$ can be parametrized by use of the generators $\lambda_i$ of $SU_4$, which are tabulated in \cite{jakobczyk}.
Using this parametrization we write
\begin{equation}
\rho=\frac{1}{4}+\frac{1}{2}\sum\limits_{i=1}^{15}m_i(t)\lambda_i, \quad  V''(t)=\frac{1}{2}\sum\limits_{i=1}^{15}\beta_i(t)\lambda_i. \label{par}
\end{equation}
%
The motivation behind the similarity transform, (\ref{sim}), is to obtain the special initial state
\begin{equation}
\rho_0=S^{-1}\rho'_0S=\frac{1}{4}-\frac{1}{2}\sqrt{\frac{3}{2}}\lambda_{15}=
\left( \begin{array}{cccc}
0&0&0&0\\
0&0&0&0\\
0&0&0&0\\
0&0&0&1\\ \end{array} \right), \label{rhopar}
\end{equation}
where all the components, $m_i$, of the Bloch vector expansion vanish, except for the component
$m_{15}=-\sqrt{3/2}$. The factor $\sqrt{3/2}$ origins from the normalization condition for this particular parametrization of the density matrix $\rho$, given by 
$\sum_{i=1}^{15}\left|m_i\right|^2=3/2$.
The transform to the coordinate frame given by (\ref{rhopar}) greatly simplifies the later analysis of the equations of motion for $\rho$. This is the geometric equivalent to the conventional choice of coordinate system for the $2$-level Bloch sphere
where the initial state is parallel to the z-axis.

\subsection{Equations of motion on the Bloch sphere}
The time evolution of the density matrix in the interaction picture is given by the
standard von Neumann equation, 
\begin{eqnarray}
\frac{d\rho(t)}{dt}&=-i[V''(t),\rho(t)].
\end{eqnarray}
By use of the Bloch vector parametrization of $\rho(t)$ and $V''(t)$, given by (\ref{par}),
we are now able to calculate the time evolution of $\rho(t)$ in a way that simplifies further
analysis of the decoherence.
The time evolution of the coefficients $m(t)$ of (\ref{par}) can be derived from the commutator relation
\begin{eqnarray}
[\lambda_i,\lambda_j]&=2if_{ijk}\lambda_k,
\end{eqnarray} 
where $f_{ijk}$ are the structure factors of $SU_4$ (a table can be found in \cite{hu}).
We obtain the following equations for the components:
\begin{eqnarray}
\dot{m}_i(t)=f_{ijk}\beta_j(t)m_k(t).
\label{timeevo}
\end{eqnarray} 

Due to the similarity transform by $S$, the initial state is always specified by $m_i=0$ for $i<15$ and $m_{15}=-\sqrt{3/2}$.
With this choice we can simplify (\ref{timeevo}) by writing $m_{i}(t)=m_{i}(0)+\alpha_{i}(t)$ and 
$$\dot{m}_i(t)=f_{ij15}\beta_j(t)[m_{15}(0)+\alpha_{15}(t)]+f_{ijk}\beta_j(t)\alpha_{k}(t)$$ 
for $k\neq 15$. To the first order in the parameter $\xi(t)$ we obtain the approximation
\begin{eqnarray}
\dot{\alpha}_i(t)\approx f_{ij15}\beta_j(t)m_{15}(0).
\label{timeevo2}
\end{eqnarray}
This approximation is valid for times sufficiently short, such that all components of the Bloch vector tangent to the Bloch sphere due to noise are small compared to the component along the initial state.

For each individual realization of the stochastic process $\xi_j(t)$, the two-qubit system evolves as a pure state on the surface of its Bloch sphere.
Since the set of density matrices corresponding to pure states is only a $6$-dimensional subset of the $14$-dimensional $4$-level Bloch sphere, there can only be six nonvanishing components in the equations of motion, (\ref{timeevo2}).
Averaged over the noise process $\xi_j(t)$, we obtain diffusive dynamics of the Bloch vector in this $6$-dimensional tangent plane.
The probability distribution for the Bloch vector will grow in width in each of the six directions, and in general the distribution width might grow faster in some directions.

Proceeding, as long as we consider times sufficiently short so that the uncertainty of the Bloch vector due to noise is small relative to the size of the Bloch sphere, we can calculate the decay of the component  $m_{15}(t)=-\sqrt{3/2}+\alpha_{15}(t)$ along the $\lambda_{15}$-axis, the  by use of the normalization condition 
\begin{eqnarray}
\alpha_{15}(t)&=\sqrt{\frac{3}{2}}-\sqrt{\frac{3}{2}-\sum\limits_{i=1}^{14}\alpha_i^{2}(t)}
\approx \sqrt{\frac{1}{6}}\sum\limits_{i=1}^{14}\alpha_i^{2}(t).
\end{eqnarray}
Averaged over initial conditions and realizations of the noise process we obtain
\begin{eqnarray}
\left\langle\alpha_{15}(t)\right\rangle&\approx\sqrt{\frac{1}{6}}\sum\limits_{i=1}^{14}\left\langle\alpha_i^{2}(t)\right\rangle,
\label{pyt}
\end{eqnarray}
where the mean square of the components $\alpha_i$ is defined by
\begin{eqnarray}
\left\langle\alpha_i^2(t)\right\rangle=\int\limits_0^t\int\limits_0^t dt_1dt_2 \, \dot{\alpha}_i(t_1)\dot{\alpha}_i(t_2).
\label{msq}
\end{eqnarray}

\subsection{Simple picture of decoherence of a two-qubit system by a single TLS}

\begin{figure}[htb]
\begin{center}
  \includegraphics[width=0.6\columnwidth,height=2in]{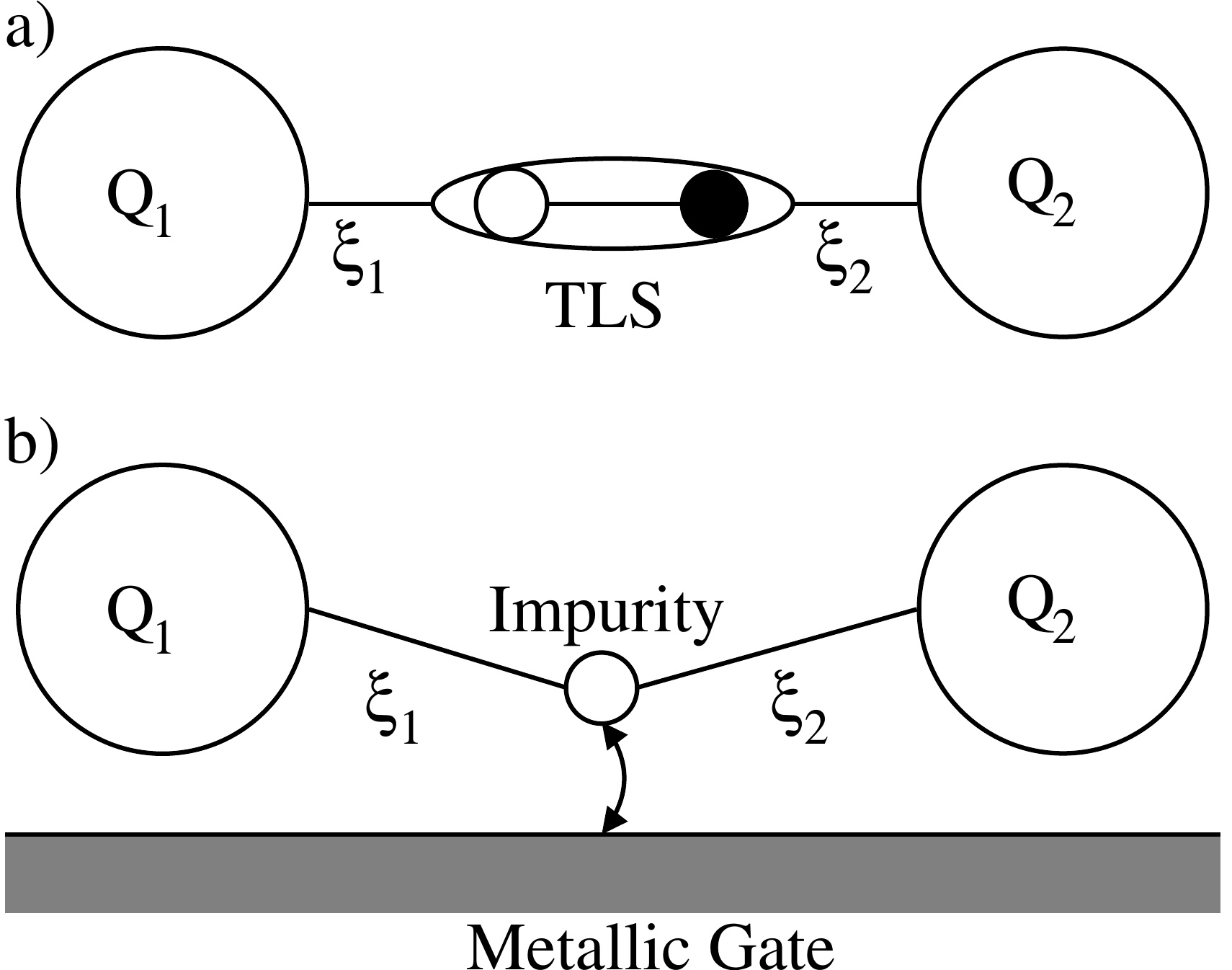}
\end{center}
 \caption{a) Two qubits, $Q_1$ and $Q_2$, coupled to a TLS in the substrate between the two qubits, where the charge can tunnel between two sites. b) The qubits are coupled to a charged impurity through its image charge on the metallic gate, the charge can tunnel between the gate and the impurity.
The coupling strengths are given by $\xi_1$ and $\xi_2$. The configuration demonstrated in a) give origin to anticorrelated noise,
while configuration b) give origin to correlated noise.}
  \label{2qubits}
\end{figure}

Before we move on to the calculations it may be instructive to look at a simple model that still contains essential features regarding the sensitivity of
our two-qubit system to external noise.
Assume two (coupled) qubits, $Q_1$ and $Q_2$, 
both coupled to an environment containing only a single quantum object, e.g., a TLS, through two channels $\xi_1$ and $\xi_2$ as illustrated in Fig.~\ref{2qubits} a) and described by the interaction Hamiltonian
\begin{equation}
H_I=(\xi_1\sigma_z^1+\xi_2\sigma_z^2)\hat{E}=\hat{X}\hat{E}.
\end{equation}
Here $\hat{E}$ is an operator acting on the quantum object.
The evolution of the state $\left|\phi_{E}(t)\right\rangle$ of the environmental quantum object is in general dependent on the state of the two-qubit system.
The decoherence of the two-qubit system is determined by tracing over the degrees of freedom of the quantum object, i.e., by the overlap 
matrix elements $\langle\phi_{E}^i(t)|\phi_{E}^j(t)\rangle$. Here $i,j$ denote the four orthogonal states of the two-qubit system. In other words, the decoherence is determined by the difference in the conditioned evolution of the state $\left|\phi_{E}^i(t)\right\rangle$ depending on the initial state of the two-qubit system.

In the configuration described in figure \ref{2qubits} b), charge fluctuations between the impurity and the metallic gate give origin to a correlated noise on the two qubits. If each qubit couple with the same strength to the gate, $\xi_1=\xi_2$, and the two qubits are prepared in a superposition of the states $\left|01\right\rangle$ and $\left|10\right\rangle$  then the expectation value of operator $\hat{X}$ coupling to the environment will vanish. The environmental quantum object is therefore unable to distinguish between these two states and therefore no entanglement build-up can take place between the environment and the qubits. In this model,  the states  $\left|01\right\rangle$ and $\left|10\right\rangle$ span a decoherence-free subspace (DFS) in the presence of correlated noise \cite{zanardi,ekert,lidar}. However, if the coupling parameters $\xi_1$ and $\xi_2$ are not exactly the same the states $\left|01\right\rangle$ and $\left|10\right\rangle$ will 
induce different potentials on the environment such that entanglement can build up. In the extreme case, where $\xi_1=-\xi_2$, which could in fact be realized by a charge trap placed between two charge qubits, as could be realized in a configuration similar to figure \ref{2qubits} a), 
completely correlated noise will be twice as efficient as the same noise from two uncorrelated sources.
Thus this possible situation needs to be accounted for in the design process of many qubit cirquits.

Furthermore, if the two-qubit system has dynamics given by an intrinsic Hamiltonian $H_0$, initial states that are not eigenstate of $H_0$ will in general
evolve in time. In this case, we will have decoherence even if the noise is completely correlated, $\xi_1=\xi_2$, and the initial state is $\left|01\right\rangle$ due to the fact that the state will decompose in eigenstates which in general might not be protected from correlated noise.

\section{Initial decoherence in the stationary path approximation}
\label{initial}
In this section, we will study the initial decoherence of our two-qubit system
focusing on the sensitivity 
of the decoherence to correlations in the noise.
Previously it has been found that the sensitivity to noise in two-qubit systems is reduced if the noise is correlated, and
the two-qubit system is prepared to be in a state within the SWAP subspace, spanned by the states $\{\left|10\right\rangle,\left|01\right\rangle\}$ \cite{averin}.
In order to check this, we calculate the decoherence rate from four different initial states, two 
of which belongs to the SWAP subspace, and two belonging to the subspace spanned by $\left|00\right\rangle$ and $\left|11\right\rangle$.

As an example to illustrate the method we will calculate the initial decoherence rate when the two-qubit system is initially prepared in the singlet state $\left|\psi_0\right\rangle=\left|10\right\rangle-\left|01\right\rangle$, where we have omitted the normalization factor.
Proceeding as described in section \ref{model}, we transform to the coordinate frame where the initial Bloch vector of the two-qubit system lies on the south pole of the Bloch sphere, $m_i=-\sqrt{3/2}$ for $i=15$ and $0$ otherwise.
The transformation matrix is for this initial state given by 
\begin{equation}
S=\frac{1}{\sqrt{2}}\left( \begin{array}{cccc}
\sqrt{2}&0&0&0\\
0&-1&0&-1\\
0&-1&0&1\\
0&0&\sqrt{2}&0\\ \end{array} \right), \quad S^{-1}=S^{T}\, .
\end{equation}
%
We can then compute the noise matrix 
$$V''=S^{T}V'S=\frac{1}{2}\sum\limits_{i=1}^{15}\beta_i\lambda_i$$
 and use (\ref{timeevo2}) to compute
the coefficients $\alpha(t)$ parameterizing the density matrix $\rho(t)$.
For the initial state $\left|\psi_0\right\rangle=\left|10\right\rangle-\left|01\right\rangle$ and the Hamiltonian given by (\ref{Ham}), the differential equations for 
$\alpha_i$
are given by Eq.~(\ref{diffs}). 
We note that there are only six nonvanishing components $\alpha_i$. 

Since the general formulas given by (\ref{diffs}) are rather complicated, it is instructive to investigate the simplified case where the two-qubit system
is at the co-resonant point $\Delta_1=\Delta_2=\Delta$. The problem simplifies further if we assume that the inter-qubit coupling is much less than the single qubit tunneling element, $\nu\ll\Delta$.
With this symmetric Hamiltonian, we get a simplified set of parameters 
$\gamma_\pm=1/\sqrt{2}$, $\eta_\pm=0$.

Furthermore, if we are interested only in the initial decoherence rate, we 
may assume the stationary path approximation, where the noise sources are stationary and decoherence is solely due to the different realizations of initial values of the noise sources at the start of each experiment (e.g., a statistical distribution of the gate charge at the start of each run of the measurement protocol \cite{paladino_08}). 
After averaging 
over many experiments with different initial conditions for the noise sources we obtain an uncertainty in the precise value of the Bloch vector of our two-qubit system.
By use of (\ref{msq}), inserting the expressions for the derivatives of the component $\alpha_i$ given in (\ref{diffs}) we obtain the following expressions 
for the mean square deviations along each component due to noise:
\begin{eqnarray}
\left\langle\alpha_9^2(t)\right\rangle&=&\left\langle\alpha_{13}^2(t)\right\rangle=
(1/8)\left\langle (\delta \xi)_- ^2\right\rangle \left[c_-(t)+c_+(t)\right]^2,\nonumber \\
\left\langle\alpha_9^2(t)\right\rangle&=&\left\langle\alpha_{13}^2(t)\right\rangle=
(1/8)\left\langle (\delta \xi)_- ^2\right\rangle \left[s_-(t)-s_+(t)\right]^2,\nonumber \\
\left\langle\alpha_{11}^2(t)\right\rangle&=&(1/4) \left\langle (\delta \xi)_- ^2\right\rangle \left[c_-(t)-c_+(t)\right]^2, \nonumber\\
\left\langle\alpha_{11}^2(t)\right\rangle&=&(1/4) \left\langle (\delta \xi)_- ^2\right\rangle \left[s_-(t)+s_+(t)\right]^2
\label{components}
\end{eqnarray}
where $(\delta \xi)_\mp \equiv\xi_1(0) \mp\xi_2(0)$, $\omega_{\pm}=\Omega_+\pm\Omega_-$,
 $s_\pm(t)=\omega_\pm^{-1}(1-\cos \omega_\pm t)$,  $s_\pm(t)=\omega_\pm^{-1} \sin \omega_\pm t$.
The expression is valid for times shorter than the correlation time of the noise sources, 
$t\ll \tau_c$.

From the normalization condition, (\ref{pyt}), and the expressions for the mean square fluctuations along the six components, (\ref{components}), we obtain the following simple formula determining the initial decoherence of the two-qubit system:
\begin{equation}
\left\langle\alpha_{15}(t)\right\rangle=\frac{\left\langle (\delta \xi)_- ^2\right\rangle }{\sqrt{6}} f(t), \quad
f(t) \equiv\frac{ c_-(t)}{\omega_-}+\frac{c_+(t)}{\omega_+} .
\end{equation}
As expected, we find that at the symmetric co-resonance point the coherence of the initial singlet state $\left|10\right\rangle-\left|01\right\rangle$ is very sensitive to the degree of correlation in the noise sources $\xi_1$ and $\xi_2$. If the noise sources
are completely correlated there is no decoherence.

The expressions for  $\langle\alpha_{15}(t)\rangle$ for different initial states are summarized in the Table~\ref{tab1}.
\begin{table}[h]
\caption{Time dependences of the coefficient $\left\langle\alpha_{15} \right\rangle$ for different initial conditions. \label{tab1}}
\begin{center}
\begin{tabular}{|c|c|}
\hline
$\left|\psi_0\right\rangle$&$\left\langle\alpha_{15}(t)\right\rangle\cdot\sqrt{6}$\\
\hline
$\left|01\right\rangle-\left|10\right\rangle$&$\langle  (\delta \xi)_-^2\rangle f(t) $\\ 
\hline
$\left|01\right\rangle+\left|10\right\rangle$&$\left\langle (\delta \xi)_-^2\right\rangle [f(t)+g(t)]
+\left\langle (\delta \xi)_+^2\right\rangle [f(t)-g(t)]$\\ \hline
$\left|00\right\rangle+\left|11\right\rangle$&$\left\langle (\delta \xi)_-^2\right\rangle [f(t)-g(t)]
+\left\langle(\delta \xi)_+^2\right\rangle [f(t)+g(t)]$\\ \hline
$\left|00\right\rangle-\left|11\right\rangle$&$\langle(\delta \xi)_+^2\rangle f(t) $\\ 
\hline
\end{tabular}
\end{center}
\end{table}
In this table,
\begin{equation}
g(t)=(\omega_+\omega_-)^{-1}(1-\cos{\omega_-t}-\cos{\omega_+t}+\cos{2\Omega t}).
\end{equation}
In the absence of correlation 
$$\langle  (\delta \xi)_- ^2\rangle=\langle (\delta \xi)_+^2\rangle=2\left(\langle [\xi_1(0)]^2\rangle + \langle [\xi_2(0)]^2\rangle \right),$$
and for the fully correlated noise $\langle  (\delta \xi)_-^2 \rangle =0$, while for the anti-correlated noise $\langle(\delta \xi)_+^2 \rangle =0$.

We find that 
the effect of correlated noise is strongly dependent on the initial state of the qubit.
The singlet state $\left|\psi\right\rangle=\left|01\right\rangle-\left|10\right\rangle$ is a decoherence-free state in the
presence of completely correlated noise, while the state $\left|01\right\rangle+\left|10\right\rangle$ is only partly protected
from correlated noise. The difference is due to the fact that the singlet state is an eigenstate of $H_0$ while the the initial state $\left|01\right\rangle+\left|10\right\rangle$ is not an eigenstate of $H_0$ and will therefore due to the time evolution obtain components in subspaces of the Hilbert space that are sensitive to correlated noise. For a general initial state, $\left|\phi_0\right\rangle$, the component in the subspace of the Hilbert space spanned by the singlet state, $(\left\langle 01\right|-\left\langle 10\right|)|\phi_0\rangle$,  will not decay in time. For this state we will find an initial decay of the coherence, and then persisting coherence of amplitude given by the overlap element.
The initial states $\left|\psi_0\right\rangle=\left|00\right\rangle-\left|11\right\rangle$ and $\left|\psi_0\right\rangle=\left|00\right\rangle+\left|11\right\rangle$ are decoherence-free and weakly protected, respectively, to anticorrelated noise.
It is important to notice that we have assumed co-resonance. If the Hamiltonian, (\ref{Hhh}), is not perfectly symmetric the symmetric states will no longer be eigenstates of $H_0$ and will therefore be less sensitive to correlated/anticorrelated noise.

To compare with the above result, if the two-qubit Hamiltonian commutes with the operator representing coupling to the noise source, i.e., for the conventional Hamiltonian \cite{ekert} $$H=\sum\limits_{j=1,2}\Delta\sigma_z^j+\nu\sigma_z^1\sigma_z^2\label{H0}+\sum\limits_{j=1,2}\xi_j(t)\sigma_z^j\, ,$$ then the subspace spanned by the set of states ${\left|01\right\rangle,\left|10\right\rangle}$ is a $2$-dimensional DFS with respect to a completely correlated noise, but vulnerable to an anticorrelated noise, and vice versa for the set of states spanned by ${\left|00\right\rangle,\left|11\right\rangle}$.

\section{Decoherence for intermediate times in the weak coupling limit}
\label{long}
In the previous section we studied the initial dynamics of the two-qubit system,
due to noise from the channels $\xi_1(t)$ and $\xi_2(t)$, by use of the stationary path approximation.
In this section we will consider the decoherence for times longer than the correlation time $\tau_c$ of the noise sources.

We proceed by use of the four-level Bloch-vector method applied in the same manner as presented earlier, by (\ref{timeevo}), (\ref{pyt}) and (\ref{msq}).
The decoherence is restricted to a six dimensional subspace of the $14$-dimensional Bloch-sphere as long as we observe the qubits for times $t\left|\xi\right|\ll 1$.
We are required to solve the integrals, of the form given by (\ref{msq}), determining the mean square component in each of the six directions
spanning the subspace.

To compare with the results of the references \cite{averin,paladino_08}, we start with the initial state $\left|\psi_0\right\rangle=\left|01\right\rangle$.
After transforming to the coordinate frame where $\rho_0=1/4-\sqrt{3/8}\lambda_{15}$,
we obtain the following equations of motion
\begin{eqnarray}
\dot{\alpha}_9(t)&=&\xi_1(t)(\lambda_1\sin{\omega_-t}+\lambda_2\sin{\omega_+t}), \nonumber\\
\dot{\alpha}_{10}(t)&=&\xi_1(t)(\mu_{1-}+\mu_{2-})(\cos{\omega_-t}-\cos{\omega_-t}),\nonumber\\
\dot{\alpha}_{11}(t)&=&[\xi_1(t)-\xi_2(t)] \left[-(\mu_{2+}+\mu_{1+})\sin{\omega_-t}\right.\nonumber\\
&&\left.+(\mu_{2+}-\mu_{1+})\sin{\omega_+t}\right],\nonumber\\
\dot{\alpha}_{12}(t)&=&0, \\
\dot{\alpha}_{13}(t)&=&\xi_2(t)(-\lambda_2\sin{\omega_-t}-\lambda_1\sin{\omega_+t}),\nonumber\\
\dot{\alpha}_{14}(t)&=&\xi_2(t)(\mu_{1-}-\mu_{2-})(-\cos{\omega_-t}+\cos{\omega_+t})
\nonumber
\label{diffs01}
\end{eqnarray}
where we have introduced the notation
\begin{eqnarray*}
\lambda_1=\eta_-^2\gamma_-^2-\eta_+^2\gamma_+^2,&\quad&\lambda_2=\eta_-^2\gamma_+^2-\eta_+^2\gamma_-^2,\nonumber\\
\mu_{1\pm}=\eta_+\eta_-(\gamma_+^2\pm\gamma_-^2), &&\mu_{2\pm}=\gamma_+\gamma_-(\eta_-^2\pm\eta_+^2).
\end{eqnarray*}

When computing the mean square of each component $\langle\alpha_i^2(t)\rangle$, by use of \eref{msq},
we assume time translation invariance, which allows us to make the
standard transformation \cite{girvinrmp}
\begin{equation}
\left\langle\alpha_i^2(t)\right\rangle
=\int\limits_0^tdT\int\limits_{-\infty}^{\infty}d\tau\dot{\alpha}_i\left(-\frac{\tau}{2}+T\right)\dot{\alpha}_i\left(\frac{\tau}{2}+T\right),
\label{transformint}
\end{equation}
where $\tau=t_2-t_1$ and $T=(t_1+t_2)/2$.
We have also assumed observation times to be longer than the correlation time of the noise sources, $t\gg\tau_c$, which allows us to extend the integral over $\tau$ to infinity.
In addition, we have to assume that the coupling to the environment is sufficiently weak to ensure that
the six components of the initial ``hypersurface''  where dynamics take place, $\langle\alpha_i^2(t)\rangle$ for $i=9-14$,  are small compared to $\langle m_{15}(t)\rangle$ .

Inserting (\ref{diffs01}) into (\ref{transformint}), we obtain terms in the integrand proportional to
$\cos (\Omega_+\pm\Omega_-)\tau$, $\cos{2(\Omega_+\pm\Omega_-)T}$
and $\cos (\pm \Omega_\pm \tau\pm 2\Omega_\pm T)$.
  If we assume that $t\gg 1/\omega_+$, the terms that oscillate as a function of $T$ will be negligible compared to 
the slowly varying ones.
By this approximation the expressions for the mean square of the six components, from the initial state $\left|\psi_0\right\rangle=\left|01\right\rangle$, reduce to the following:
\begin{eqnarray}
\langle\alpha_9^2(t)\rangle &\approx\int\limits_0^t \! \! \frac{dT}{2} \! \!\int\limits_{-\infty}^{\infty}\! \! d\tau \langle\xi_1(\tau)\xi_1(0) \rangle [\lambda_1^2 \cos \omega_-\tau
+\lambda_2^2\cos \omega_+\tau].
\end{eqnarray}
Here we omitted the explicit expressions for the other five components. We notice that the integral grows linearly in time.

By use of the normalization condition, (\ref{pyt}), we can make the following approximation for $t\left|\xi_{1,2}\right|\ll 1$,
\begin{equation}
\langle\alpha_{15}(t)\rangle \approx (1/\sqrt{6})\sum\limits_{i=1}^{15}\langle\alpha_i^{2}(t)\rangle 
\approx t\Gamma/\sqrt{6} 
\label{rate}
\end {equation}
where we introduced the decoherence rate $\Gamma$ characterizing the decay of the Bloch vector given by $\langle\alpha_{15}(t)\rangle$ to lowest order in $t$.
With the initial condition $\left|\psi_0\right\rangle=\left|01\right\rangle$ we obtain the following expression for $\Gamma$:
\begin{eqnarray}
&& \Gamma=(1/\sqrt{6})\Big\{
\sum_{\pm} S_{11}(\omega_\mp)+ S_{22}(\omega_\pm)]a_\pm
\nonumber \\ 
&& \quad -2S_{12}(\omega_-)(\mu_{2+}+\mu_{1+})^2
- 2S_{12}(\omega_+)(\mu_{2+}-\mu_{1+})^2\Big\}, \nonumber \\
&&a_+=\lambda_1^2+(\mu_{2+}+\mu_{1+})^2+(\mu_{2-}+\mu_{1-})^2,  \nonumber \\
&&a_-=\lambda_2^2+(\mu_{2+}-\mu_{1+})^2+(\mu_{2-}+\mu_{1-})^2\, .
\label{rate01}
\end{eqnarray}
Here we have introduced the spectral density 
\begin{eqnarray}
S_{ij}(\omega)=\int\limits_{-\infty}^{\infty}\! \! d\tau\langle\xi_i(\tau)\xi_j(0)\rangle \cos(\omega \tau)
\end{eqnarray}
 for the Fourier transform of the (cross) correlation spectra from channels $i,j$ at frequency $\omega$. The expression given by (\ref{rate01}) is valid for $\max\{1/\omega_+,\tau_c\}\ll t\ll 1/\left|\xi_{1,2}\right|$. We note that the noise at the two frequencies, $\Omega_+\pm\Omega_-$, is the only relevant component of the noise in the weak coupling limit. This result agrees with the findings of \cite{averin}. 
It is, however, important to stress that the above result is to first order in the coupling strength $\xi_i$.
For a single qubit, with energy splitting $\Delta$, subject to transverse noise, it is known \cite{shnirman04,zbergli06} that we will get a contribution to the dephasing from the noise at zero frequency
$$
\langle\alpha_i^2(t)\rangle\propto \xi^2S(\omega)+\xi^4S(0)+..,
$$
where $\omega=\Delta$ is the qubit frequency at zero bias, and the zero frequency contribution is due to the net shift in the precession rate due to the noise, $\omega=\sqrt{\Delta^2+\xi^2}=\Delta+\xi^2/2\Delta^2$.
This effect will also be present in the multi-qubit problem.

\begin{figure}[t]
\centering
\includegraphics[width=0.49\columnwidth,height=1.5in]{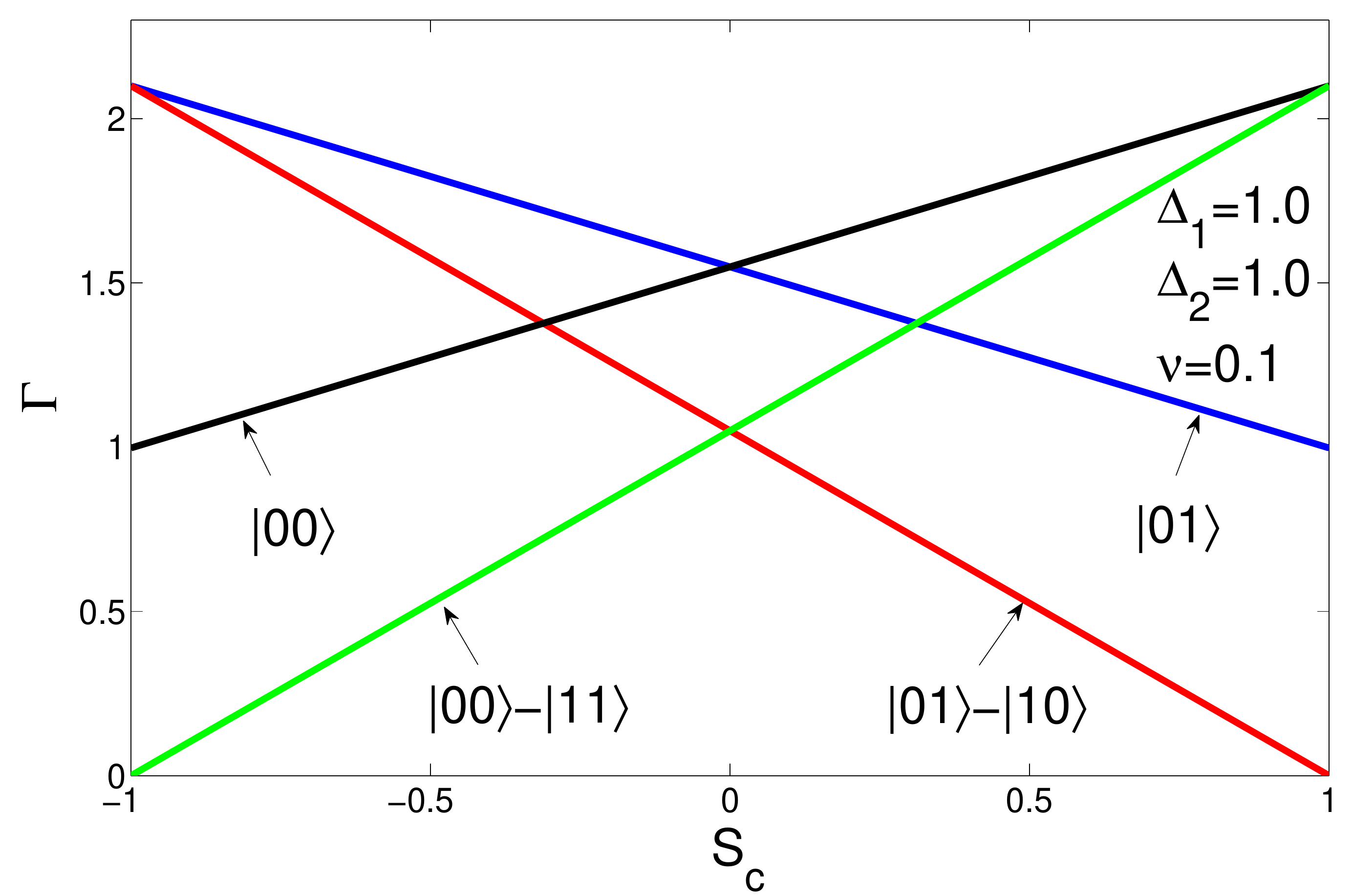} 
\includegraphics[width=0.49\columnwidth,height=1.5in]{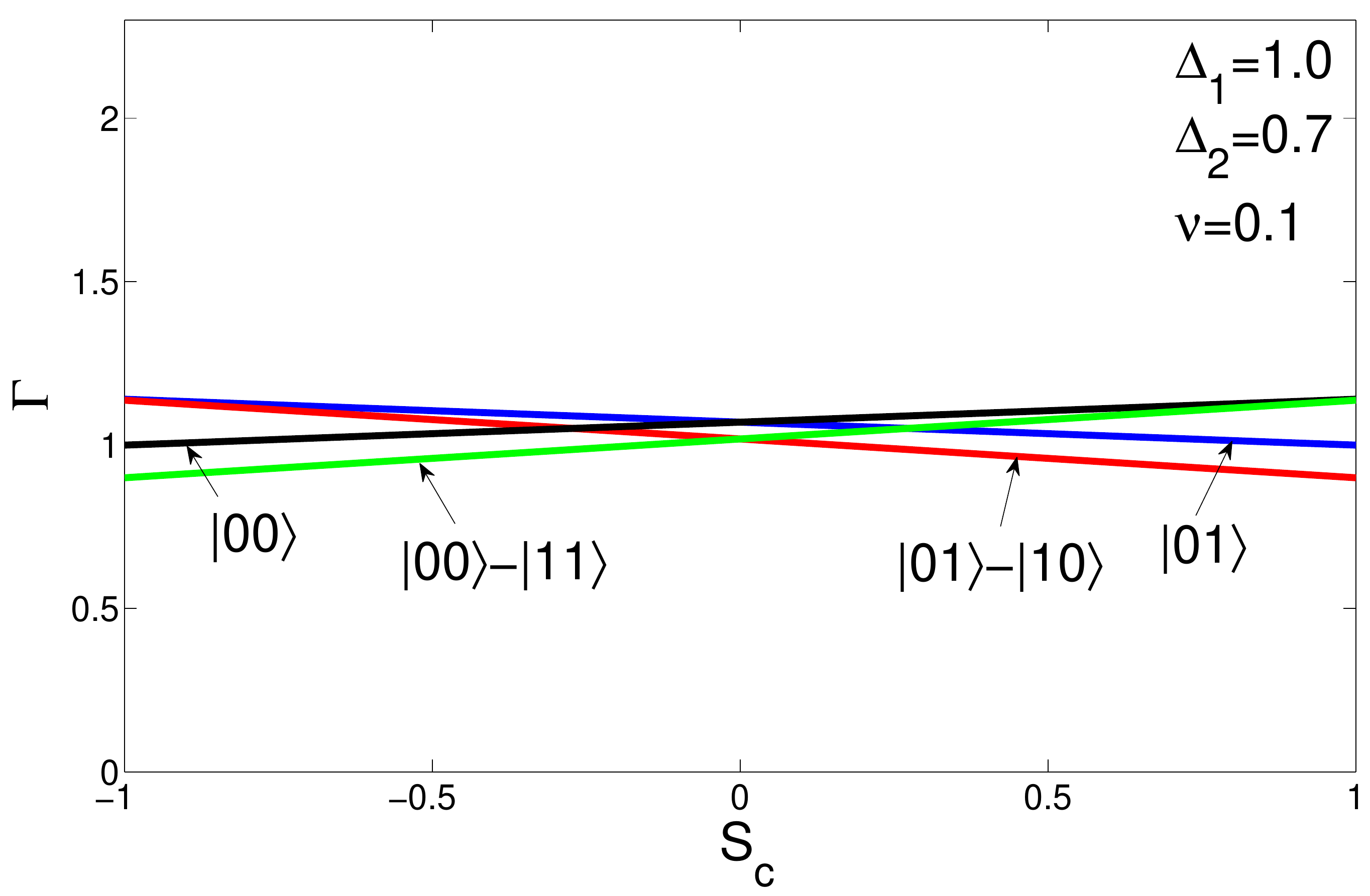} 
  \caption{(Color online) Sensitivity of the decoherence rate $\Gamma$ to correlations between $\xi_1(t)$ and $\xi_2(t)$ and to symmetry in the Hamiltonian $H_0$. We plot the total decoherence rate $\Gamma$ against the correlation coefficient $S_C$ for four different initial states, $\left|\psi_0\right\rangle=
\left|01\right\rangle-\left|10\right\rangle$ (red), $\left|\psi_0\right\rangle=\left|01\right\rangle$ (blue), $\left|\psi_0\right\rangle=\left|00\right\rangle$ (black) and $\left|\psi_0\right\rangle=\left|00\right\rangle-\left|11\right\rangle$ (green).
In the top figure we are at the co-resonance point, where $\Delta_1=\Delta_2$, while in the bottom figure $\Delta_1=0.7\Delta_2$.
The coupling parameter is $\nu=0.1$. }
  \label{corrsens}
\end{figure}

In the work by Averin and Rabenstein \cite{averin}, it was also reported that the decoherence rate was more sensitive to correlation when the two-qubit system was prepared in the state $\left|\psi_0\right\rangle=\left|01\right\rangle$ than in the state $\left|\psi_0\right\rangle=\left|00\right\rangle$.
Contrary to these results, we find that the two states are initially equally sensitive to correlations. If the noise sources have the same magnitude, $S_{11}(\omega)=S_{22}(\omega)$, we can express the decoherence rate as a function of the correlation parameter $S_C$ given by (\ref{corrcoeff}). The rate as a function of correlation is plotted in figure \ref{corrsens}, 
where we find that the coherence time of the state $\left|\psi_0\right\rangle=\left|01\right\rangle$ is enhanced in the presence of 
correlated noise, while the coherence time of the initial state $\left|\psi_0\right\rangle=\left|00\right\rangle$ is diminished by the same factor.
Our results is fully consistent with those of Averin, and the ``seeming'' discrepancy has the following explanation. 
In reference \cite{averin}, the oscillations of the
probability $p_{1,2}(t)$ to find $Q_1$ and $Q_2$ in the excited state at time $t$
are plotted.
Our results for the decoherence rate agree at short times (in the regime $t\left|\xi\right|\ll 1$ where our approximation is valid), but thereafter
in \cite{averin}
slowly damped oscillations of amplitude $\sim 0.7$ for the initial state $\left|01\right\rangle$ were found. 
This behavior is due to the fact that the state $\left|01\right\rangle$ is not an eigenstate of $H_0$. For the Hamiltonian (\ref{Hhh}) at the co-resonance point, the eigenstates are the
usual singlet and triplet states. The singlet state is protected against correlated noise, while the triplet is not. The initial state $\left|01\right\rangle$ has equal components in both subspaces. In the presence of a correlated noise, the component in the triplet will therefore decay rapidly while we will find persisting oscillations due to the finite component in the singlet subspace.
The Hamiltonian 
considered in \cite{averin} is not fully
symmetric, but the picture is still valid. One component 
decays at a fast rate,
while the component in the approximate ``singlet''-subspace decays much slower.
Interestingly, it was reported in \cite{weiss}, that concurrence oscillations from the initial state $\left|01\right\rangle$ lasts even longer than expected from its components in the DFS and its compliment, due to an interference effect between the components that decreases the decoherence. 

From a practical point of view, however, we can focus on the initial decoherence rate. If the purity of the state has decayed more than a few percent quantum computation by error correction is practically impossible.
Furthermore, our results show that the initial state is even more important than it was previously reported \cite{averin,paladino_08}.
In the presence of completely correlated noise, $S_C=1$, the singlet state $\left|\psi_0\right\rangle=\left|01\right\rangle-\left|10\right\rangle$
is decoherence-free. However, it is important to mention that this is the only decoherence-free state. There is no $2$-dimensional DFS for our Hamiltonian $H_0$. The state $\left|\psi_0\right\rangle=\left|01\right\rangle-\left|10\right\rangle$ is decoherence-free
\textit{if and only if} we are at the co-resonance point, as shown in figure \ref{corrsens}. If we move away from this point the sensitivity to correlations is reduced. This behavior is controlled by the symmetry of the Hamiltonian. For our Hamiltonian,
(\ref{Hhh}), the relevant parameters are 
$\kappa_1=\Delta_-/\Delta_+$ and $\kappa_2=\Delta_+/\nu$. 
For either $\kappa_1\ll 1$ or $\kappa_2\ll 1$ the Hamiltonian has a high degree of symmetry and there exist states that are protected against correlated or anticorrelated noise. If the Hamiltonian has low degree of symmetry, initial states that are initially symmetric or antisymmetric with respect to the noise source, i.e., the state $\left|\psi_0\right\rangle=\left|01\right\rangle-\left|10\right\rangle$, will decompose in the non-symmetric eigenstates of $H_0$ and quickly lose its symmetry by the time evolution induced by $H_0$. In this case the sensitivity to correlations in the noise source will be very weak.

\section{Discussion}
\label{discussion}
Our Bloch-vector treatment of the two-qubit decoherence problems shows that the two qubit problem can be separated in two regimes. In the symmetric regime, the symmetry of the initial state $\left|\psi_0\right\rangle$ is conserved by the Hamiltonian in the absence of noise. In this regime, the system is very sensitive to correlations or anticorrelations in the noise sources $\xi_1(t)$ and $\xi_2(t)$ acting on qubits $Q_1$ and $Q_2$, respectively. This sensitivity, which is consistent with the general DFS picture, 
is dependent on the initial state and on the parameters of the problem in a more complex way than it was reported in \cite{paladino_08,averin}, where it was assumed that states in the SWAP subspace spanned by the states $\left|01\right\rangle$ and $\left|10\right\rangle$ should be less influenced by a correlated noise. 
At the same time,
the decoherence of states outside this subspace should only be weakly influenced by correlations. This thought is based on the concept of
DFSs defined as the states where the the dissipative part of the Markovian master equation is zero \cite{karasik}, but does not take fully into account the intrinsic dynamics of the system. A general theory for dynamically stable DFSs was developed in \cite{karasik}.
We find that it is both the symmetry of the initial state and how much this state overlaps with an eigenstate of the Hamiltonian in the absence of noise that determines the rate of decoherence. 
For the Hamiltonian given by (\ref{Ham}), at the co-resonance point, all the three initial states $\left|01\right\rangle,\left|10\right\rangle$ and $\left|01\right\rangle-\left|10\right\rangle$ look similar initially for an external noise source coupled diagonally in the basis-forming states, see figure \ref{2qubits}. However, only the last state is an eigenstate of $H_0$ and therefore fully protected during the time evolution, i.e., a true DFS.

While previous works, except for the general classifications of DFSs, have mainly focused on a correlated noise in the SWAP subspace, we find that this is not the only interesting subspace in this regard.
Considering two qubits at the co-resonance point
we find that the subspace spanned by the states $\left|00\right\rangle,\left|11\right\rangle$ is protected against an antisymmetric noise. For this system, an antisymmetric noise also leads to increase of the initial decoherence rate $\Gamma$ for states in the SWAP subspace, and an increase in $\Gamma$ by  factor 2 for the singlet state.

For the case of noise due to charged traps in the amorphous substrate, a trap located 
between two qubits, giving rise to anticorrelated noise, seems just as plausible as the opposite situation of strict correlations.
In an experimental setup, one might take advantage of this situation by first looking for correlated or anticorrelated noise, and thereafter decide the encoding of the two qubits in the subspace that is less sensitive to the noise present in the relevant setup. 

Our treatment, by use of the $4$-level Bloch vector construction, gives a tractable set of equations compared to alternative methods.
The familiar geometrical concepts used when treating two-level systems by the same method can be carried over to higher dimensions.
In our treatment, different rates of decoherence along different directions on the Bloch sphere, as well as their sensitivity to correlations in the noise, follow naturally from the formalism. One may hope that this method might be extended also to even higher level systems, e.g., to multi-qubit systems relevant for the analysis of the noise in a quantum computer.
While we are able to derive simple analytical expressions for the two-qubit system, the three-qubit system will require a $24$ dimensional Bloch-sphere, and obtaining analytical expressions in this case seems cumbersome except in very special limits, due to the diagonalization of the system Hamiltonian $H_0$ and of $\rho$, which are both $8\times 8$ matrices.
The method is still suitable for numerical impementation, where one might take advantage of the familiar geometric concepts from the two-level problem.

In conclusion, we have demonstrated a generalized Bloch-sphere method by treating the problem of two coupled qubits coupled to 
statistically-independent or correlated noise sources.
By use of this method we find expressions for the dephasing of the two-qubit system as a function of the degree of correlation of the noise sources. At the co-resonance point, the decoherence rate is very sensitive to the initial two-qubit state. In the absence of symmetry of the Hamiltonian, correlation in the noise seems to have very weak impact on the decoherence rate.

\appendix
\section{Detailed expressions}
\label{expressions}
\subsection{The noise matrix $V'(t)$}
The explicit expression for the matrix elements of $V'(t)=U^{-1}H_1(t)U$ is given by:
\begin{eqnarray} \label{Vrot}
V'_{11}&=&\xi_1[(\eta_-\gamma_-+\eta_+\gamma_+)^2\cos{\omega_-t}+(\eta_-\gamma_+-\eta_+\gamma_-)^2\cos{\omega_+t}]
\nonumber \\ &&
+\xi_2[(\eta_-\gamma_++\eta_+\gamma_-)^2 \cos{\omega_-t}+(\eta_-\gamma_--\eta_+\gamma_+)^2\cos{\omega_+t}],\nonumber\\
V'_{12}&=&\xi_2[\left(-\mu_{1-}+\mu_{2-}\right)\cos{\omega_-t} +\left(\mu_{1-}-\mu_{2-}\right)\cos{\omega_+t}
-i\lambda_2\sin{\omega_-t}-i\lambda_1\sin{\omega_-t}], \nonumber\\
V'_{13}&=&\xi_1[\left(\mu_{1-}+\mu_{2-}\right)\cos{\omega_-t} -\left(\mu_{1-}+\mu_{2-}\right)\cos{\omega_+t}
-i\lambda_1\sin{\omega_-t}-i\lambda_2 \sin{\omega_+t}],  \nonumber\\
V'_{14}&=&(\xi_1+\xi_2)[-i\left(\mu_{2+}+\mu_{1+}\right) \sin{\omega_-t}
+i\left(\mu_{2+}-\mu_{1+}\right)\sin{\omega_+t}], \nonumber\\
V'_{22}&=&\xi_1[(\eta_-\gamma_-+\eta_+\gamma_+)^2\cos{\omega_-t}+(\eta_-\gamma_+-\eta_+\gamma_-)^2\cos{\omega_+t}]
\nonumber \\ &&
-\xi_2[(\eta_-\gamma_++\eta_+\gamma_-)^2\cos{\omega_-t}+(\eta_-\gamma_--\eta_+\gamma_+)^2\cos{\omega_+t}],\nonumber\\
V'_{23}&=&(\xi_1-\xi_2)[i \left(\mu_{2+}+\mu_{1+}\right) \sin{\omega_-t}
-i \left(\mu_{2+}-\mu_{1+}\right) \sin{\omega_+t}],\nonumber\\
V'_{24}&=&\xi_1[-\cos{\omega_-t}\left(\mu_{1-}+\mu_{2-}\right) +\cos{\omega_+t}\left(\mu_{1-}+\mu_{2-}\right)
-i\sin{\omega_-t}\lambda_1-i\sin{\omega_+t}\lambda_2],\nonumber\\
V'_{33}&=&-\xi_1[(\eta_-\gamma_-+\eta_+\gamma_+)^2\cos{\omega_-t}+(\eta_-\gamma_+-\eta_+\gamma_-)^2\cos{\omega_+t}]
\nonumber \\ &&
+\xi_2[(\eta_-\gamma_++\eta_+\gamma_-)^2\cos{\omega_-t}+(\eta_-\gamma_--\eta_+\gamma_+)^2\cos{\omega_+t}],\nonumber\\
V'_{34}&=&\xi_2[\left(\mu_{1-}-\mu_{2-}\right)\cos{\omega_-t} -\left(\mu_{1-}-\mu_{2-}\right)\cos{\omega_+t}
-i \lambda_2 \sin{\omega_-t}-i \lambda_1 \sin{\omega_+t}],\nonumber\\
V'_{44}&=&-\xi_1[(\eta_-\gamma_-+\eta_+\gamma_+)^2\cos{\omega_-t}+(\eta_-\gamma_+-\eta_+\gamma_-)^2\cos{\omega_+t}]
\nonumber \\ &&
-\xi_2[(\eta_-\gamma_++\eta_+\gamma_-)^2\cos{\omega_-t}+(\eta_-\gamma_--\eta_+\gamma_+)^2\cos{\omega_+t}], 
\end{eqnarray}
where we have only given the upper triangle
of the Hermitian matrix $V'$.

\subsection{Differential equations for the density matrix components $\alpha_i$}
The differential equations for the components of the density matrix evolving from the initial state $\left|\psi_0\right\rangle=\left|10\right\rangle-\left|01\right\rangle$:
\begin{eqnarray}
\dot{\alpha}_9&=&\frac{\xi_1(t)}{\sqrt{2}}\big[\lambda_1\sin{\omega_-t}+\lambda_2\sin{\omega_+t}\big]
-\frac{\xi_2(t)}{\sqrt{2}}\big[\lambda_2\sin{\omega_-t}+\lambda_1\sin{\omega_+t}\big],\nonumber\\
\dot{\alpha}_{10}&=&\frac{\xi_1(t)}{\sqrt{2}}\big[-(\mu_{1-}+\mu_{2-})\cos{\omega_-t}
+(\mu_{1-}+\mu_{2-})\cos{\omega_+t}\big]\nonumber\\
&&+\frac{\xi_2(t)}{\sqrt{2}}\big[-(\mu_{1-}-\mu_{2-})\cos{\omega_-t}
+(\mu_{1-}-\mu_{2-})\cos{\omega_+t}\big],\nonumber\\
\dot{\alpha}_{11}&=&(\xi_1(t)-\xi_2(t))\big[(\mu_{1+}+\mu_{2+})\sin{\omega_-t}
+(\mu_{1+}-\mu_{2+})\sin{\omega_+t}\big],\nonumber\\
\dot{\alpha}_{12}&=&-\frac{\xi_1(t)}{\sqrt{2}}\big[(\eta_-\gamma_-+\eta_+\gamma_-)^2\cos{\omega_-t}+(\eta_-\gamma_+-\eta_+\gamma_-)^2\cos{\omega_+t}\big]
\nonumber\\ &&
+\frac{\xi_2(t)}{\sqrt{2}}\big[(\eta_-\gamma_++\eta_+\gamma_-)^2\cos{\omega_-t}+(\eta_-\gamma_--\eta_+\gamma_-)^2\cos{\omega_+t}\big],\nonumber\\
\dot{\alpha}_{13}&=&\frac{\xi_1(t)}{\sqrt{2}}\big[\lambda_1\sin{\omega_-t}+\lambda_2\sin{\omega_+t}\big]
-\frac{\xi_2(t)}{\sqrt{2}}\big[\lambda_2\sin{\omega_-t}+\lambda_1\sin{\omega_+t}\big],\nonumber\\
\dot{\alpha}_{14}&=&\frac{\xi_1(t)}{\sqrt{2}}\big[-(\mu_{1-}+\mu_{2-})\cos{\omega_-t}
+(\mu_{1-}+\mu_{2-})\cos{\omega_+t}\big]\nonumber\\
&&+\frac{\xi_2(t)}{\sqrt{2}}\big[-(\mu_{1-}-\mu_{2-})\cos{\omega_-t}
+(\mu_{1-}-\mu_{2-})\cos{\omega_+t}\big]. 
\label{diffs}
\end{eqnarray}

\providecommand{\newblock}{}


\begin{thebibliography}{10}
\expandafter\ifx\csname url\endcsname\relax
  \def\url#1{{\tt #1}}\fi
\expandafter\ifx\csname urlprefix\endcsname\relax\def\urlprefix{URL }\fi
\providecommand{\eprint}[2][]{\url{#2}}

\bibitem{clarke08}
Clarke J and Wilhelm F~K 2008 {\em Nature\/} {\bf 453}(7198) 1031

\bibitem{gaebel_nphys}
Gaebel T, MDomhan, Popa I, Wittmann C, Neumann P, Jelezko F, Rabeau J~R,
  Stavrias N, Greentree A~D, Prawer S, Meijer J, Twamley J, Hemmer P~R and
  Wrachtrup J 2006 {\em Nature Physics\/} {\bf 2} 408--413

\bibitem{Hanson_science}
Hanson R, Dobrovitski V~V, Feiguin A~E, Gywat O and Awschalom D~D 2008 {\em
  Science\/} {\bf 320} 352--355

\bibitem{iontraps}
Häffner H, Roos C and Blatt R 2008 {\em Physics Reports\/} {\bf 469} 155 --
  203

\bibitem{ladd}
Ladd T~D, Press D, De~Greve K, McMahon P~L, Friess B, Schneider C, Kamp M,
  H\"ofling S, Forchel A and Yamamoto Y 2010 {\em Phys. Rev. Lett.\/} {\bf
  105}(10) 107401

\bibitem{hanson_rmp}
Hanson R, Kouwenhoven L~P, Petta J~R, Tarucha S and Vandersypen L~M~K 2007 {\em
  Rev. Mod. Phys.\/} {\bf 79}(4) 1217--1265

\bibitem{coish}
Coish W~A and Loss D 2004 {\em Phys. Rev. B\/} {\bf 70}(19) 195340

\bibitem{tsyplyatyev}
Tsyplyatyev O and Loss D 2011 {\em Phys. Rev. Lett.\/} {\bf 106}(10) 106803

\bibitem{martinis04}
Simmonds R~W, Lang K~M, Hite D~A, Nam S, Pappas D~P and Martinis J~M 2004 {\em
  Phys. Rev. Lett.\/} {\bf 93} 077003

\bibitem{martinis05}
Martinis J~M, Cooper K~B, McDermott R, Steffen M, Ansmann M, Osborn K~D, Cicak
  K, Oh S, Pappas D~P, Simmonds R~W and Yu C~C 2005 {\em Phys. Rev. Lett.\/}
  {\bf 95} 210503

\bibitem{martinis10}
Shalibo Y, Rofe Y, Shwa D, Zeides F, Neeley M, Martinis J~M and Katz N 2010
  {\em Phys. Rev. Lett.\/} {\bf 105} 177001

\bibitem{tian}
Tian L and Simmonds R~W 2007 {\em Phys. Rev. Lett.\/} {\bf 99} 137002

\bibitem{zgalperinprl}
Galperin Y~M, Altshuler B~L, Bergli J and Shantsev D~V 2006 {\em Phys. Rev.
  Lett.\/} {\bf 96} 097009

\bibitem{shnirman05}
Shnirman A, Sch\"on G, Martin I and Makhlin Y 2005 {\em Phys. Rev. Lett.\/}
  {\bf 94} 127002

\bibitem{zgalperinmeso}
Galperin Y~M, Altshuler B~L and Shantsev D~V 2004 Low-frequency noise as a
  source of dephasing of a qubit {\em Fundamental Problems of Mesoscopic
  Physics\/} ({\em NATO Science Series\/} vol 154) (Springer Netherlands) pp
  141--165

\bibitem{bergli}
Bergli J, Galperin Y~M and Altshuler B~L 2009 {\em New Journal of Physics\/}
  {\bf 11} 025002

\bibitem{tsai_nature}
Pashkin Y~A, Yamamoto T, Astafiev O, Nakamura Y, Averin D~V and Tsai J~S 2003
  {\em Nature\/} {\bf 421} 823--826

\bibitem{dicarlo}
DiCarlo L, Reed M~D, Sun L, Johnson B~R, Chow J~M, Gambetta J~M, Frunzio L,
  Girvin S~M, Devoret M~H and Schoelkopf R~J 2010 {\em Nature\/} {\bf 467}
  574--578

\bibitem{martinis_nature}
Martinis J~M and et~al 2010 {\em Nature\/} {\bf 467} 570--573

\bibitem{governale}
Governale M, Grifoni M and Schon G 2001 {\em Chemical Physics\/} {\bf 268}
  273--283

\bibitem{wilhelm}
Storcz M~J and Wilhelm F~K 2003 {\em Phys. Rev. A\/} {\bf 67}(4) 042319

\bibitem{averin}
Rabenstein K and Averin D~V 2003 {\em Turk J Physics\/} {\bf 27} 313

\bibitem{nori}
You J~Q, Hu X and Nori F 2005 {\em Phys. Rev. B\/} {\bf 72}(14) 144529

\bibitem{guo}
Hu Y, Zhou Z~W, Cai J~M and Guo G~C 2007 {\em Phys. Rev. A\/} {\bf 75}(5)
  052327

\bibitem{paladino_08}
D'Arrigo A, Mastellone A, Paladino E and Falci G 2008 {\em New Journal of
  Physics\/} {\bf 10} 115006

\bibitem{faoro}
Faoro L and Hekking F~W~J 2010 {\em Phys. Rev. B\/} {\bf 81}(5) 052505

\bibitem{mastellone}
Mastellone A, D'Arrigo A, Paladino E and Falci G 2008 {\em The European
  Physical Journal - Special Topics\/} {\bf 160}(1) 291--300

\bibitem{zanardi}
Zanardi P and Rasetti M 1997 {\em Phys. Rev. Lett.\/} {\bf 79}(17) 3306--3309

\bibitem{lidar}
Lidar D~A, Chuang I~L and Whaley K~B 1998 {\em Phys. Rev. Lett.\/} {\bf 81}(12)
  2594--2597

\bibitem{fortunato02}
Fortunato E~M, Viola L, Hodges J, Teklemariam G and Cory D~G 2002 {\em New
  Journal of Physics\/} {\bf 4} 5

\bibitem{fortunato}
Fortunato E~M, Viola L, Pravia M~A, Knill E, Laflamme R, Havel T~F and Cory D~G
  2003 {\em Phys. Rev. A\/} {\bf 67}(6) 062303

\bibitem{karasik}
Karasik R~I, Marzlin K~P, Sanders B~C and Whaley K~B 2008 {\em Phys. Rev. A\/}
  {\bf 77}(5) 052301

\bibitem{lidarbook}
Lidar D~A and B W 2003 Decoherence-free subspaces and subsystems {\em
  "Irreversible Quantum Dynamics", F. Benatti and R. Floreanini (Eds.)\/} vol
  622 (Springer Lecture Notes in Physics, Berlin) pp 83--120

\bibitem{goyal}
Goyal S~K, Simon B~N, Singh R and Simon S 2011 Geometry of the generalized
  bloch sphere for qutrit

\bibitem{siennicki}
Jakobczyk L and Siennicki M 2001 {\em Physics Letters A\/} {\bf 286} 383 -- 390

\bibitem{kimura}
Kimura G 2003 {\em Physics Letters A\/} {\bf 314} 339 -- 349

\bibitem{jakobczyk}
Jakobczyk L and Siennicki M 2001 {\em Physics Letters A\/} {\bf 286} 383 -- 390

\bibitem{hu}
Hu B 1974 {\em Phys. Rev. D\/} {\bf 9}(6) 1825--1834

\bibitem{ekert}
Palma G~M, Suominen K~A and Ekert A~K 1996 {\em Proc. Roy. Soc. Lond.\/} {\bf
  A452} 567--584

\bibitem{girvinrmp}
Clerk A~A, Devoret M~H, Girvin S~M, Marquardt F and Schoelkopf R~J 2010 {\em
  Rev. Mod. Phys.\/} {\bf 82} 1155--1208

\bibitem{shnirman04}
Makhlin Y and Shnirman A 2004 {\em Phys. Rev. Lett.\/} {\bf 92}(17) 178301

\bibitem{zbergli06}
Bergli J, Galperin Y~M and Altshuler B~L 2006 {\em Phys. Rev. B\/} {\bf 74}(2)
  024509

\bibitem{weiss}
Campagnano G, Hamma A and Weiss U 2010 {\em Physics Letters A\/} {\bf 374} 416
  -- 423

\end{thebibliography}
\end{document}